\documentclass{article}
\usepackage{latexsym}
\begin{document}


%
%

\title{IS IT PHYSICALLY SOUND  TO ADD A  TOPOLOGICALLY MASSIVE TERM TO
THREE-DIMENSIONAL MASSIVE ELECTROMAGNETIC OR GRAVITATIONAL MODELS
? }

\author{\footnotesize ANTONIO ACCIOLY and MARCO DIAS}

\date{}
\maketitle

\noindent
Instituto de F\'{\i}sica Te\'orica,
Universidade Estadual Paulista,
01405-900 S\~ao Paulo, SP, Brazil;
e-mail: accioly@ift.unesp.br.


\begin{abstract}
The addition of a topologically massive term to an admittedly
non-unitary three-dimensional massive model, be it an
electromagnetic system or a gravitational one, does not cure its
non-unitarity. What about the enlargement of avowedly unitary
massive models by way of a topologically massive term? The
electromagnetic models remain unitary after the topological
augmentation  but, surprisingly enough, the gravitational ones
have their unitarity spoiled. Here we analyze these issues and
present the explanation why unitary massive gravitational models,
unlike unitary massive electromagnetic ones, cannot coexist from
the  viewpoint of unitarity with topologically massive terms. We
also discuss the novel features of the three-term effective field
models that are  gauge-invariant.

 \it{Topologically massive models; massive electromagnetic models; massive gravitational models;
 unitarity; effective field models.}
\end{abstract}


\section{Introduction}
In the last two decades much attention has been devoted to the
study of the remarkable properties of gauge theories in (2 + 1)
dimensions.\cite{1,2} Certainly, it would not be an exaggeration
to claim that by now these properties are  not only
well-appreciated but also well-understood. Therefore, it should be
natural, at least from a naive point of view, to expect that the
addition of a Chern-Simons term to massive electromagnetic or
gravitational models would produce systems endowed with properties
that, in principle, should be as exciting as those concerning the
well-known theories of Maxwell-Chern-Simons or
Einstein-Chern-Simons. Our aim here is to analyze these massive,
topologically massive, models. Since, currently, there are two
distinct non-topological mass-generating mechanisms for gauge
fields: adding the well-known Proca/Fierz-Pauli, or the more
sophisticated higher-derivative electromagnetic/higher-derivative
gravitational, terms, our analysis will comprise topologically
massive Proca electromagnetism (TMPE), topologically massive
higher-derivative electromagnetism (TMHDE), which is also known as
Podolsky-Chern-Simons planar electromagnetism,\cite{3}
topologically massive Fierz-Pauli gravity (TMFPG), and
topologically massive higher-derivative gravity (TMHDG). These
systems will be examined for both possible sign choices of the
Maxwell/Einstein Lagrangian, as well as in its absence, which
implies  that they are the most general such models. Both TMPE and
TMFPG are not  gauge-invariant due the presence of an explicit
mass term, but the three-term models with higher-derivatives are
gauge-invariant.

 We are particularly interested in two issues that
are somewhat correlated:
\begin{itemize}
\item[i] The compatibility---from the point of view of the
unitarity---between massive electromagnetic or gravitational
models and topologically massive terms.

\item[ii] The exciting physics resulting  from the utilization  of the
 gauge-invariant three-term systems as effective field models. We
remark that it was recently shown that boson-boson bound states do
exist in the framework of three-dimensional higher-derivative
electromagnetism augmented by a topological Chern-Simons
term.\cite{3}
\end{itemize}

To probe the  unitarity of the massive, topologically massive,
models, we  will make use of an uncomplicated and easily handling
algorithm that converts the task of checking the unitarity, which
in general demands much work, into a straightforward algebraic
exercise.\cite{4} The prescription consists basically in
saturating the propagator with external conserved currents,
compatible with the symmetries of the system, and in examining
afterwards the residues of the saturated propagator ($SP$) at each
of their simple poles. We use natural units throughout.

\section{Massive, Topologically Massive, Electromagnetic Models}
The Lagrangian for TMPE is the sum of Maxwell, standard Proca
mass, Chern-Simons, terms, namely,

\begin{equation}
{\mathcal L_{\mathrm{TMPE}}} = -a \frac{F_{\mu\nu} F^{\mu\nu}}{4}
+ \frac{1}{2} m^2 A^\mu A_\mu + \frac{s}{2}
\varepsilon_{\mu\nu\rho} A^\mu \partial^\nu A^\rho,
\end{equation}

\noindent while the Lagrangian for TMHDE is the sum of Maxwell,
higher-derivative,\cite{5} gauge-fixing (Lorentz-gauge), and
Chern-Simons, terms, {\it i.e.},

\begin{equation}
{\mathcal L_{\mathrm{TMHDE}}} = -a \frac{F_{\mu\nu} F^{\mu\nu}}{4}
+ \frac{l^2}{2} \partial_\nu F^{\mu\nu}\partial^\lambda
F_{\mu\lambda} - \frac{1}{2\lambda}(\partial_\nu A^\nu)^2 +
\frac{s}{2} \varepsilon_{\mu\nu\rho} A^\mu
\partial^\nu A^\rho.
\end{equation}

\noindent Here, $F_{\mu\nu} = \partial_\nu A_\mu - \partial_\mu
A_\nu$ is the usual electromagnetic tensor field, $l$ is a cutoff,
$s> 0$ is the topological mass,
 and $a$ is a convenient parameter
that can take the values $+1$ (Maxwell's term with the standard
sign), $-1$ (Maxwell's term with the ``wrong sign"), or $0$
(absence of the Maxwell's term). The corresponding propagators are
given by

\begin{equation}
\mathcal{P}_{\mathrm{TMPE}} = - \frac{a k^2 - m^2}{(a k^2- m^2)^2
- s^2 k^2}\; \theta + \frac{1}{m^2} \;\omega - \frac{s}{(a k^2 -
m^2)^2 - s^2 k^2} \;S,
\end{equation}

\begin{equation}
\mathcal{P}_{\mathrm{TMHDE}} = \frac{l^2 k^4 - a k^2}{(l ^2 k ^4
-ak^2)^2 - s^2 k^2} \;\theta - \frac{\lambda}{k^2} \;\omega -
\frac{s}{(l ^2 k ^4 -ak^2)^2 - s^2 k^2}\;S,
\end{equation}

\noindent where $\theta_{\mu\nu} \equiv \eta_{\mu\nu} -
\frac{\partial_\mu
\partial_\nu}{\Box}$ and $\omega_{\mu\nu} = \frac{\partial_\mu
\partial_\nu}{\Box}$ are, respectively, the usual transverse and
longitudinal vector projector operators, $S_{\mu\nu} \equiv
\varepsilon_{\mu\rho\nu}  \partial^\rho$ is the operator
associated with the topological term, and $\eta_{\mu\nu} $ is the
Minkowski metric. Our signature conventions are $(+,-,-)$,
$\varepsilon^{012}=+1=\varepsilon_{012}$. Now, the algorithm from
Ref. 4 says that all we have to do in order to check the unitarity
of a massive, topologically massive, electromagnetic model is to
verify whether the residues at each simple pole of the
$\theta$-component of the propagator in the basis $\{\theta,
\omega, S\}$ which, for short, we will designate as $f_\theta$,
are $\leq 0$. We use this recipe in the following to test the
unitarity of TMPE and TMHDE, in this order.

\subsection{ Checking the unitarity of TMPE}
We start our analysis  by setting  the parameter $a$  in  Eq. (1)
equal to $+1$ because it must be positive both in the Proca
($s=0$) and Maxwell-Chern-Simons  ($m=0$) limits{\cite 2}. Now,
the $\theta$-component of the propagator in the basis $\{\theta,
\omega, S \}$ is,  according to Eq. (3), equal to $f_\theta =
\frac{m^2 -k^2}{(k^2 - m^2_+)(k^2 -m^2_-)}$, where $m_\pm =
\frac{1}{2}\left[ \sqrt{4m^2 + s^2}  \pm s \right]$. Therefore,
the model has two degrees of freedom with masses $m_+$ and $m_-$,
which is precisely what Deser and Tekin have found using a rather
different approach.\cite{6} Our result is also in agreement with
another works existing in the literature.\cite{7} On the other
hand, it is trivial to show that both $\mathrm{Res}\; f_\theta
|_{\;k^2 =m^2_+}$ and $\mathrm{Res}\; f_\theta |_{\;k^2 =m^2_-}$
are less than zero. Thence, TMPE with the Maxwell's term with the
usual sign is  unitary. Choosing $a=-1$, we see that if $s^2
>4m^2$, then $f_\theta = \frac{m^2 + k^2}{ (k^2 - m^2_+)(k^2 -
m^2_-)}$, with $m_\pm = \frac{1}{2} \left[ s \pm \sqrt{s^2 - 4m^2}
\right]$. A straightforward calculation allows us  to conclude
that $\mathrm{Res}\; f_\theta |_{\;k^2 =m^2_+} >0$, and
$\mathrm{Res}\; f_\theta |_{\;k^2 =m^2_-} <0$, implying that, if
$s^2 >4m^2$, TMPE with Maxwell's term with the wrong sign is
non-unitary . It is worth mentioning that this system, despite
having acceptable values for the masses, faces ghost problems. Of
course, if $s^2 <4m^2$ the two roots of $x^2 + x( 2m^2 -s^2) + m^
4 =0$, where $k^2 \equiv x$, are imaginary; note also that for
$s^2=4m^2$ the two hitherto complex roots coalesce and the masses
are simply $m_+=m_-=\frac{s}{2}$: these models are never viable.
We focus, at last, on the case $a=0$ (absence of the Maxwell's
term). This model was analyzed long ago by Deser and Jackiw, who
came to the conclusion that setting $a=0$ yields just another
version of the Maxwell-Chern-Simons theory and so it is equivalent
to choosing $m^2=0$.\cite{8}

\subsubsection{Discussion}
Note that Proca electromagnetism $(s=0)$ with $a=-1$ contains
tachyons; however, TMPE with $a=-1$ and $s^2
>4 m^2$ is plagued by ghosts but not by tachyons: the particle
content of the model is one non-tachyonic spin-1 ghost of mass
$m_+$ and one massive spin-1 normal particle of mass $m_-\;$.
Thence, a field theory built from this model would not be
satisfactory from the point of view of their fundamentals. It
could regarded, perhaps, as an effective field theory, {\it i.e.},
a low-energy approximation to a more fundamental theory.
Nonetheless, the condition $s>2m$ is greatly discouraging as far
as the possibility of applying this kind of model, for instance,
to some condensed matter systems where one deals, in general, with
low-energy excitations. Interesting enough, only the model with
$a=+1$ may be viewed as physically sound. Why is this
 so? Because the aforementioned system  has a Lagrangian that reproduces the Lagrangians of
well-behaved physical models when the appropriate limits are
taken. Indeed, if $s=0$, we recover Proca electromagnetism; on the
other hand, setting $m=0$, we obtain Chern-Simons
electromagnetism. It is remarkable that we also arrive at a nice
physical model by removing the Maxwell's term: the system with
$a=0$ and the ``self-dual" model of  Ref. 8 are equivalent.

\subsection{Checking the unitarity of TMHDE}

 Based on the above informations, we have every reason to
begin the  unitarity analysis of TMHDE by setting $a=+1$ in Eq.
(2). The calculations are  now more complicated because, unlike
the previous model, this represents in general three
 massive excitations rather than two massive ones. Since the
 $\theta$-component of the propagator concerning TMHDE with $a=+1$ can
 be written as $f_\theta=\frac{M^2(x-M^2)}{x^3 -2M^2x^2 + M^4x -
 M^4s^2}$, where $M\equiv \frac{1}{l}$,  we have to
 analyze the nature, as well as the signs, of the roots of the cubic equation $x^3
 + a_2x^2 + a_1x + a_0=0$, where $a_2 \equiv -2M^2, a_1 \equiv M^4,$ and $a_0
  \equiv -M^4 s^2$. Taking into account that we are only interested in
  those roots that are both real and unequal, we require
  $D<0$, where $D \equiv Q^3 + R^2$, with $Q$ and $R$ being, in
  this order, equal to $\frac{3a_1 - a^2_2}{9}$ and
  $\frac{9a_1a_2 -27a_0 -2a^3_2}{54}$, is the polynomial
  discriminant. Performing the computations we get $D=M^8s^2
  \left[ \frac{s^2}{4} - \frac{M^2}{27} \right]$, implying that
  only and if only $s^2 <  \frac{4M^2}{27}$ will the roots be real
  and unequal. Our next step is to verify whether or not these roots
  are positive. This can be accomplished by building the
  Routh-Hurwitz array,\cite{9} namely,

\begin{table}[h]
\begin{center}
\begin{tabular}{cc}
1&$M^4$\\ $-2M^2$&$-M^4s^2$\\ $M^2\left(M^2
-\frac{s^2}{2}\right)$&0\\ $-M^4s^2$&0
\end{tabular}
\end{center}
\end{table}

\noindent Noting that there are three signs changes in the first
column of the array above, we conclude that all the three  roots
are positive. In summary, if $s^2 < \frac{4m^2}{27}$, TMHDE with
$a=+1$ is a model with acceptable values for the masses. Denoting
these roots as $x_1, x_2$, and $x_3$, and assuming without any
loss of generality that $x_1 > x_2 > x_3$, we get

\begin{eqnarray}
f_\theta  &=& \frac{M^2(x_1 -M^2)}{(x_1 -x_2)(x_1 - x_3)}
\frac{1}{x -x_1} + \frac{M^2(x_2 -M^2)}{(x_2 -x_1)(x_2 - x_3)}
\frac{1}{x -x_2} \nonumber \\ &&+ \frac{M^2(x_3 -M^2)}{(x_3
-x_1)(x_3 - x_2)} \frac{1}{x -x_3}. \nonumber
\end{eqnarray}

\noindent Hence, TMHDE with $a=+1$ will be unitary if the
conditions $ x_1 -M^2<0, \; x_2 -M^2>0,\;$ and $x_3 -M^2<0$ hold
simultaneously. Obviously, this will never occur, which allows us
to conclude that TMHDE with the Maxwell's term with the standard
sign is non-unitary. If $a=-1$, $\;  f_\theta =
\frac{M^2(x+M^2)}{x^3 + 2M^2x^2 + M^4x -M^4s^2}\;$. Since the
polynomial discriminant, $D=M^8s^2 \left[ \frac{s^2}{4} +
\frac{M^2}{27} \right]$,  for the cubic equation $x^3 + 2M^2x^2 +
M^4x - M^4s^2=0$ is greater than zero, one of the roots of the
equation is real and the other two are complex conjugates, which
means that the system with $a=-1$ is forbidden. To finish our
analysis we set $a=0$ in Eq. (2). In this case $f_\theta =
\frac{xM^2}{x^3 - M^4s^2}$, and the polynomial discriminant
related to $x^3 - M^4s^2 =0$ is greater than zero. This model, as
the previous one, is also forbidden.

\subsubsection{Discussion}
Should we expect intuitively that TMHDE with $a=+1$ faced unitary
problems? The answer is affirmative. In fact, setting $s=0$, for
instance, in its Lagrangian, we recover the Lagrangian for the
usual Podolsky electromagnetism which is non-unitary. Nonetheless,
Podolsky-Chern-Simons (PCS) planar electromagnetism with $a=+1$
and $s^2 < \frac{4M^2}{27}$, despite being haunted by ghosts, has
normal massive modes. Since the existence of these well-behaved
excitations is subordinated to the condition
 $s< \frac{2M}{\sqrt{27}}$, we are really encouraged to regard
 this system as an effective field model. It is quite remarkable
 that the coupling of PCS planar electrodynamics with a
 charged scalar field, produces an attractive interaction between
 equal charge bosons. To see this we need to know beforehand the
 expression of
 the effective non-relativistic potential for the interaction of
 two charged-bosons in the center-of-mass frame. A somewhat
 involved computation, yields

\begin{eqnarray}
V(r) &=& - \frac{s Q^2}{\pi m l^4}\left[ \frac{ l^4}{s^2}
\frac{1}{r^2} + \frac{1}{r} \sum_{j} B_j {\sqrt{|x_j|}}
\;K_1({\sqrt{|x_j|}}\; r) \right]{\bf L}  \nonumber \\ &&+
\frac{Q^2}{2 \pi l^4} \left[ \sum_{j} A_j K_0({\sqrt{|x_j|}}\; r)
\right],
\end{eqnarray}

\noindent where $A_1 \equiv \frac{1 +l^2
x_1}{(x_1-x_2)(x_1-x_3)}\;,\;A_2 \equiv \frac{1 +l^2
x_2}{(x_2-x_1)(x_2-x_3)}\;, \;A_3 \equiv \frac{1 +l^2
x_3}{(x_3-x_1)(x_3-x_2)}\;,\;B_1 \equiv \frac{-(1+
l^2x_1)^2}{s^2(x_1-x_2)(x_ 1-x_3)}\; ,\;B_2 \equiv
\frac{-(1+l^2x_2)^2}{s^2(x_2-x_1)(x_2-x_3)}\;,\;$ and $\;B_3
\equiv \frac{-(1+l^2x_3)^2}{s^2(x_3-x_1)(x_3-x_2)}\;,\;$
$x_1,\;x_2,\;$ and $\;x_3$ are the roots of the equation $x^3 +
\frac{2x^2}{l^2} + \frac{x}{l^4} + \frac{s^2}{l^4} =0 \;,$ {\bf L}
is the angular momentum, $K$ is the modified Bessel function, and
$Q$ and $m$ are, respectively, the charge and the mass of the
scalar boson. On the other hand, the radial Schr\"odinger equation
associated with this potential is given by

\begin{eqnarray}
\left[\frac{d^2}{d r^2}+  \frac{1}{r} \frac{d}{d r} \right]
\mathcal {R}_{n \bar{l}} +m \left[E_{n \bar{l}} -
V^{eff}_{\bar{l}}\right] \mathcal {R}_{n \bar{l}}=0,
\end{eqnarray}

\noindent where

\begin{eqnarray}
V^{eff}_{\bar{l}}(r) &=& - \frac{s Q^2}{\pi m l^4}\left[
\frac{l^4}{s^2} \frac{1}{r^2} + \frac{1}{r} \sum_{j} B_j
{\sqrt{|x_j|}} \;K_1({\sqrt{|x_j|}}\; r) \right]\bar{l} \nonumber
\\ &&+ \frac{Q^2}{2 \pi l^4} \left[ \sum_{j} A_j
K_0({\sqrt{|x_j|}}\; r) \right] + \frac{{\bar l}^2}{m r^2}.
\nonumber
\end{eqnarray}

\noindent Here $\bar{l}$ denotes the eigenvalues of the operator
$\bf{L}$. In terms of the dimensionless parameters $y \equiv s
r,\; \alpha \equiv \frac{Q^2}{\pi s},\; b_j \equiv \frac{s^2}{l^4}
B_j,\; X_j \equiv \frac{|x_j|}{s},\; \beta \equiv \frac{m}{s},\;
a_j \equiv \frac{A_j}{l^4},$ and $\; \tilde {E}_{n \bar{l}} \equiv
\frac{mE_{n \bar{l}}}{s^2}$, Eq. (6) reads

\begin{eqnarray}
\left[\frac{d^2}{d y^2}+  \frac{1}{y} \frac{d}{d y} \right]
\mathcal {R}_{n \bar{l}} +\left[ \tilde{E}_{n \bar{l}} -
\tilde{V}^{eff}_{\bar{l}}\right] \mathcal {R}_{n \bar{l}}=0,
\nonumber
\end{eqnarray}

\noindent with

\begin{eqnarray}
\tilde{V}^{eff}_{\bar{l}} \equiv -\frac{\bar{l} (\alpha -
\bar{l})}{y^2}  +\frac{\alpha \beta}{2} \sum_{j} a_j K_0(X_j\; y)
-\frac{\alpha \bar{l}}{y} \sum_{j} b_j X_j K_1(X_j\; y). \nonumber
\end{eqnarray}

\noindent We call attention to the fact that
$\tilde{V}^{eff}_{\bar{l}}$ behaves as $\frac{\bar{l}^2}{y^2}$ at
the origin and as $\frac{\bar{l} ( \bar{l} -\alpha)}{y^2}$
asymptotically. Now, in four dimensions, the anomalous factor of
$\frac{4}{3}$ in the Abraham-Lorentz model for the electron does
not show up if $l
> \frac{1}{2}r_e$, where $r_e$ denotes the classical radius of the
electron.\cite{10} Therefore, we assume $l \ll 1$. In this limit
the derivative of the potential with respect to $y$ reduces to

\begin{eqnarray}
\frac{d}{d y} \tilde{V}^{eff}_{\bar{l}} \sim \frac{2 \bar{l}
(\alpha - \bar{l})}{y^3} -\left[\frac{2\alpha \bar{l}}{y^2} +
\frac{\alpha\beta}{2} \right] K_1(y) - \frac{\alpha \bar{l}}{y}
K_0(y). \nonumber
\end{eqnarray}

\noindent Supposing $\bar{l} >0$, without any loss of generality,
we promptly see that, if $\bar{l} >\alpha$, the potential is
strictly decreasing. The remaining possibility is $\bar{l} <
\alpha$. In this interval $\tilde{V}^{eff}_{\bar{l}}$ approaches
$+\infty$ at the origin and $0^{-}$ for $y\rightarrow+\infty$,
which is indicative of a local minimum. Consequently, the
existence of the attractive potential is subordinated to the
conditions $a \ll 1$ and $0 < \bar{l} <\alpha$. One can show that
the effective potential with $l \ll 1$ and $ 0< \bar{l} < \alpha$
can bind a pair of identical charged-scalar bosons.\cite{3}
Accordingly, the addition of the topologically
 massive term to Podolsky's electromagnetism with $a=+1$---an
 admittedly non-unitary model---did not cure its non-unitary
 problem; nonetheless, the condition for the resulting
 three-term model to be free of tachyons  gives rise to a
 constraint between the topological and Podolsky masses which is
 responsible for  a scalar attractive interaction.

\section{Massive, Topologically Massive, Gravitational Models (MTMG)}

 TMFPG is defined by the Lagrangian

\begin{eqnarray}
 {\mathcal L}_\mathrm{TMFPG} &=&  a\frac{2}{{\bar \kappa}^2} {\sqrt{g}} \; R - \frac{m^2}{2} \left( {h_{\mu\nu}}^2 -
h^2 \right) \nonumber
\\ &&+ \frac{1}{\mu}
\epsilon^{\lambda\mu\nu} {\Gamma^{\rho}}_{\lambda\sigma} \left(
\partial_\mu {\Gamma^\sigma}_{\rho\nu} +\frac{2}{3}
{\Gamma^\sigma}_{\mu\beta}{\Gamma^\beta}_{\nu\rho}\right),
\end{eqnarray}

\noindent at quadratic order in ${\bar \kappa}$, where $ {\bar
\kappa}^2$ is a suitable constant that in four dimensions is equal
to $24\pi G$, with $G$ being  Newton's constant.\cite{11} Here
$g_{\mu\nu} \equiv \eta_{\mu\nu} + {\bar \kappa} h_{\mu\nu}$, $\mu
> 0 $  is a dimensionless parameter,
 and $h \equiv \eta_{\mu\nu} h^{\mu\nu}$. Indices are raised
(lowered) with $\eta^{\mu\nu}$ ($\eta_{\mu\nu}$). The Lagrangian
related to TMHDG, in turn, is given by

\begin{eqnarray}
{\mathcal L}_\mathrm{TMHDG} &=& \sqrt{g} \left(a
\frac{2R}{\kappa^2} +\frac{\alpha}{2} R^2 +\frac{ \beta}{2}
R_{\mu\nu}^2 \right) \nonumber \\ &&+ \frac{1}{\mu}
\epsilon^{\lambda\mu\nu} {\Gamma^{\rho}}_{\lambda\sigma} \left(
\partial_\mu {\Gamma^\sigma}_{\rho\nu} +\frac{2}{3}
{\Gamma^\sigma}_{\mu\beta}{\Gamma^\beta}_{\nu\rho}\right),
\end{eqnarray}

\noindent where $\alpha$ and $\beta$ are suitable constants with
dimension $L$. For the sake of simplicity, the  gauge-fixing term
was omitted. Note that the parameter $a$ appearing in Eqs. (7) and
(8) allows for choosing the Einstein sign's term or even removing
it.

The propagator related to TMFPG is

\begin{eqnarray}
\mathcal{P}_\mathrm{TMFPG} &=& -\frac{1}{m^2} P^1 - \frac{M^2
\left( m^2 + a \Box \right)}{\Box^3 + M^2 a^2 \Box^2 + 2 a m^2 M^2
\Box + M^2 m^4} P^2 \nonumber \\ &&-\; \frac{M}{\Box^3 + M^2 a^2
\Box^2 + 2am^2M^ 2 \Box + M^2m^4 } P \nonumber \\ &&-\; \frac{m^2
+ a \Box}{2m^4} {\overline P}^{\;0} + \frac{1}{2m^2}
{\overline{\overline P}}^{\;0},
\end{eqnarray}

\noindent where $M \equiv \mu /{\bar \kappa^2}$. On the other
hand, linearizing  Eq. (8) and adding to the result the
gauge-fixing term ${\mathcal L}_{\mathrm gf} = \frac{1}{2\lambda}
({h_{\mu\nu}}^{,\;\nu} -  \frac{1}{2}h_{,\;\mu})^2$ (de Donder
gauge), we find  that the propagator concerning TMHDG takes the
form

\begin{eqnarray}
\mathcal{P}_\mathrm{TMHDG}&=&
\frac{1}{\Box[a+b(\frac{3}{2}+4c)\Box]}\overline{\overline{P}}^{\;0}
+ \frac{2\lambda}{k^2}P^1
+\;\frac{1}{\Box[a+b(\frac{3}{2}+4c)\Box]}P^0 \nonumber \\
&&+\;\frac{4M}{\Box[M^2b^2\Box^2-4(abM^2-1)\Box+4M^2a^2]}P
\nonumber
\\ &&-\;\frac{2M^2(2a-b\Box)}{\Box[M^2b^2\Box^2-4(abM^2-1)\Box+4M^2a^2]}P^2\nonumber\\
&&+\;\left[
-\frac{4\lambda}{\Box}+\frac{2}{\Box[a+b(\frac{3}{2}+4c)\Box]}\right]\overline
{P}^{\;0},
\end{eqnarray}
\noindent where $b\equiv\frac{\beta\kappa^2}{2}$, $c \equiv
\frac{\alpha}{\beta}$, and $M \equiv \frac{\mu}{\kappa^2}$.   Our
conventions are ${R^\alpha}_{\beta\gamma\delta} = -
\partial_\delta {\Gamma^\alpha}_{\beta\gamma} + ... ,\; R_{\mu\nu}
= {R^\alpha}_{\mu\nu\alpha},\; R= g^{\mu\nu} R_{\mu\nu}$, where
$g_{\mu\nu}$ is the metric tensor, and signature $(+, -, -)$. The
propagators  were calculated using the basis  $$\{P^1, P^2, P^0,
{\overline P}^{\;0}, {\overline{\overline P}}^{\;0} , P \},\;$$
where $ P^1, P^2, P^0, {\overline P}^{\;0},$ and
${\overline{\overline P}}^{\;0}$, are the usual three-dimensional
Barnes-Rivers operators,\cite{12} namely,

\begin{eqnarray}
P^1_{\mu\nu,\;\rho\sigma} = \frac{1}{2} \left( \theta_{\mu\rho}\;
\omega_{\nu\sigma} + \theta_{\mu\sigma}\;\omega_{\nu\rho} +
\theta_{\nu\rho}\;\omega_{\mu\sigma} +
\theta_{\nu\sigma}\;\omega_{\mu\rho} \right),\nonumber
\end{eqnarray}

\begin{eqnarray}
P^2_{\mu\nu,\;\rho\sigma} = \frac{1}{2} \left(
\theta_{\mu\rho}\;\theta_{\nu\sigma} +
\theta_{\mu\sigma}\;\theta_{\nu\rho} -
\theta_{\mu\nu}\;\theta_{\rho\sigma} \right), \nonumber
\end{eqnarray}

\begin{eqnarray}
P^0_{\mu\nu,\;\rho\sigma} = \frac{1}{2}
\theta_{\mu\nu}\;\theta_{\rho\sigma},\;\;{\overline
P}^{\;0}_{\mu\nu,\;\rho\sigma} =
\omega_{\mu\nu}\;\omega_{\rho\sigma}, \nonumber
\end{eqnarray}

\begin{eqnarray}
{\overline{\overline P}}^{\;0}_{\mu\nu,\;\rho\sigma} =
\theta_{\mu\nu}\;\omega_{\rho\sigma} +
\omega_{\mu\nu}\;\theta_{\rho\sigma},\nonumber
\end{eqnarray}

\noindent and $P$ is the operator associated with the linearized
Chern-Simons term, {\it i.e.},

\begin{eqnarray}
 P_{\mu\nu,\;\rho\sigma} \equiv \frac{\Box
\partial^\lambda}{4}[\epsilon_{\mu\lambda\rho}\;\theta_{\nu\sigma}
+ \epsilon_{\mu\lambda\sigma}\;\theta_{\nu\rho} +
\epsilon_{\nu\lambda\rho}\;\theta_{\mu\sigma} +
\epsilon_{\nu\lambda\sigma}\;\theta_{\mu\rho}]. \nonumber
\end{eqnarray}

\noindent According to Ref. 4, the saturated propagator concerning
the MTMG, is given by

\begin{eqnarray}
SP_\mathrm{MTMG} = \left[ T^{\mu\nu} T_{\mu\nu} - \frac{1}{2} T^2
\right] f_{P^2} + \frac{1}{2} T^2 f_{P^0},
\end{eqnarray}

\noindent where $T^{\mu\nu}$ is the external conserved current
that, obviously, is symmetric in the indices $\mu$ and $\nu$, $T
\equiv \eta_{\mu\nu} T^{\mu\nu}$, and $f_{P^2}$ and $f_{P^0}$ are,
respectively, the components $P^2$ and $P^0$ of the propagator in
the basis $\{P^1, P^2, P^0, {\overline P}^{\;0},
{\overline{\overline P}}^{\;0} , P \}\;$. Therefore, to find out
whether or not the gravitational model is unitary, we must compute
$SP_\mathrm{MTMG}$ using Eq. (11) and determine afterwards the the
residue at each simple pole of $SP_\mathrm{MTMG}$: If all the
residues are $\geq 0$, the model is unitary; however, if at least
one of them is negative, the system is non-unitary. The unitarity
analysis is greatly facilitated  if we take into account that
$\left[T^{\mu\nu} T_{\mu\nu} - \frac{1}{2} T^2 \right]_{k^2=m^2}
>0$ and $\;\left[T^{\mu\nu} T_{\mu\nu} -  T^2 \right]_{k^2=0} =0$,
where $m \geq 0$ is the mass of a generic physical particle
associated with the MTMG, and $k$ is the corresponding momentum
exchanged. Using this prescription, we check in the following the
unitarity of TMFPG and TMHDE, in this order.

\subsection{Checking the unitarity of TMFPG}

To begin with, we set $a=-1$ in Eq. (8) because we want to recover
the  Einstein-Chern-Simons Lagrangian in the $m=0\;$
limit---topologically massive gravity is a theory that requires
$a=-1$ to be ghost-free.\cite{2} The corresponding saturated
propagator is given by

\begin{eqnarray}
SP_\mathrm{TMFPG} = \left[ T^{\mu\nu} T_{\mu\nu} - \frac{1}{2} T^2
\right] \frac{M^2 (m^2 + k^2)}{k^6 - M^2 k^4 - 2m^2 M^2 k^2 - M^2
m^4}. \nonumber
\end{eqnarray}

\noindent Our next step is to study the roots of the cubic
equation $x^3 - M^2 x^2 - 2m^2 M^2 x^2 - M^2 m^4 =0$. Since the
discriminant, $D= M^4 m^6 \left[ \frac{M^2}{27} + \frac{m^2}{4}
\right]$, related to this equation is greater than zero, the model
at hand is unphysical and must be rejected. Consequently, we turn
our attention to the system with $a=+1$. Now, we have to consider
the roots of the equation $x^3 - M^2 x^2 + 2m^2 M^2 x^2 - M^2 m^4
=0$, whose polynomial discriminant can be written as  $D= M^4 m^6
\left[ \frac{m^2}{4} - \frac{M^2}{27} \right]$. Therefore, if
$\frac{m^2}{M^2} < \frac{4}{27}$, our equation has three distinct
 real roots. The corresponding Routh-Hurwitz array is

\begin{table}[h]
\begin{center}
\begin{tabular}{cc}
1&$2m^2M^2$\\ $-M^2$&$-M^2m^4$\\ $m^2\left(2M^2 -m^2\right)$&0\\
$-M^2m^4$&0
\end{tabular}
\end{center}
\end{table}

\noindent Accordingly, the system with $a=+1$ and $\frac{m^2}{M^2}
< \frac{4}{27}$ has acceptable values for the masses. Proceeding
just as we have done for TMHDE with $a=+1$ and $s^2 <
\frac{4M^2}{27}$, we promptly obtain

\begin{eqnarray}
SP_\mathrm{TMFPG} &=& \frac{F(k)(m^2 -x_1)}{(x_1 -x_2)(x_1 - x_3)}
\frac{1}{k^2 -x_1} + \frac{F(k)(m^2 -x_2)}{(x_2 -x_1)(x_2 - x_3)}
\frac{1}{k^2 -x_2} \nonumber \\ &&+ \frac{F(k)(m^2 -x_3)}{(x_3
-x_1)(x_3 - x_2)} \frac{1}{k^2 -x_3}, \nonumber
\end{eqnarray}

\noindent where $F(k) \equiv \{ T^{\mu\nu} (k) T_{\mu\nu} (k)
-\frac{1}{2} \left[ T(k)\right]^2 \}M^2$. From the above, we
clearly see that this model will be unitary if $ m^2 > x_1, \; m^2
< x_2,\;$ and $m^2 >x_3$. We thus come to the conclusion that
TMFPG with $a=+1$ and $\frac{m^2}{M^2} < \frac{4}{27}$ is
non-unitary, which means that the topological term is responsible
for breaking down the unitarity of the harmless Fierz-Pauli
gravity. If $a=0$, the discriminant associated with the equation
$x^3 - M^2 m^4 =0$ is greater than zero, which implies that this
model is physically unsound. We remark that our conclusions are in
complete agreement with those of Ref. 6 where a quite different
approach to the unitarity problem was employed.

\subsubsection{Discussion}

The above results points to an important and at the same time
interesting question: Why can unitary massive electromagnetic
models coexist in peace  with topologically massive terms, whereas
unitary massive gravitational ones cannot? The root of the problem
lies in the rather odd way Einstein-Chern-Simons theory is
constructed: The presence of the ghosts in the dynamical field is
avoided by choosing the Einstein's term with the wrong
sign.\cite{2} This is trivial to show. Indeed, writing the
Einstein-Chern-Simons Lagrangian as

\begin{eqnarray}
{\mathcal L} =  a \sqrt{g}  \frac{2R}{\kappa^2} + \frac{1}{\mu}
\epsilon^{\lambda\mu\nu} {\Gamma^{\rho}}_{\lambda\sigma} \left(
\partial_\mu {\Gamma^\sigma}_{\rho\nu} +\frac{2}{3}
{\Gamma^\sigma}_{\mu\beta}{\Gamma^\beta}_{\nu\rho}\right),
\noindent
\end{eqnarray}

\noindent with $a = \pm 1$, we promptly see that the corresponding
saturated propagator is given by

\begin{equation}
SP =  -\frac{1}{a}\left(T^{\mu\mu} T_{\mu\nu} - \frac{1}{2} T^2
\right) \frac{1}{k^2-M^2} - \frac{1}{a} \left(T^{\mu\mu}
T_{\mu\nu} - T^2 \right) \frac{1}{k^2}.
\end{equation}

\noindent Thus, to render the theory unitary we are obliged to set
$a=-1$ in Eq. (13). Note that as far as these three-term systems
are concerned, we are always in a dilemma: Which value should we
assign to $a$, $-1$ or $+1$? If we single out $a=-1$, for
instance,   we recover
 Einstein-Chern-Simons theory when the non-topological massive
 term is removed; however, in the absence of the topological term ,we do not get  a nice physical
 theory because now the Einstein's term has the wrong sign. On the other hand, if we
 pick out $a=+1$, we do not recover Einstein-Chern-Simons theory
 when the non-topological massive term is removed. In other words,
 due to the unusual Einstein sign's term in the Lagrangian
 concerning Einstein-Chern-Simons theory, the augmented systems do
 not reduce to well-behaved physical models in the suitable
 limits. Note that these idiosyncrasies do not occur in the framework of  massive, topologically
 massive, electromagnetic models because the Maxwell sign's term
 concerning Maxwell-Chern-Simons theory is the same as that of the
 usual Maxwell's theory.

\subsection{Checking the unitarity of TMHDG}

Assuming $ a\neq 0$, the $SP$ concerning TMHDG can be written as

\begin{eqnarray}
SP_\mathrm{TMHDG}&=&\frac{M^2b}{2} \frac{
(T^{\mu\nu}T_{\mu\nu}-\frac{1}{2}T^2)}{ k^2 -M^2_1}
\frac{-1+\sqrt{1-2abM^2}}{\sqrt{1-2abM^2} \left[1 - abM^2
-\sqrt{1-2abM^2} \right] }\nonumber\\ &&+ \frac{M^2b}{2} \frac{
(T^{\mu\nu}T_{\mu\nu}-\frac{1}{2}T^2)}{ k^2 -M^2_2}
\frac{1+\sqrt{1-2abM^2}}{\sqrt{1-2abM^2} \left[1 - abM^2
+\sqrt{1-2abM^2} \right] }\nonumber\\
&&+\;\frac{T^{\mu\nu}T_{\mu\nu}-T^2}{ak^2}+\frac{\frac{1}{2}T^2}{a(k^2-m^2)},
\end{eqnarray}

\noindent where
\begin{eqnarray*}
M_1^2&\equiv&
\left(\frac{2}{b^2M^2}\right)[1-abM^2-\sqrt{1-2abM^2}],\nonumber\\
M_2^2&\equiv&
\left(\frac{2}{b^2M^2}\right)[1-abM^2+\sqrt{1-2abM^2}],\nonumber\\
m^2&\equiv& \frac{a}{b(3/2+4c )}.
\end{eqnarray*}

\noindent It is interesting to note that $M^2_1 \rightarrow M^2$,
and $M^2_2 \rightarrow +\infty\;$, as $b \rightarrow 0$, which
implies that Eq. (14) reduces to Eq. (13) when $\alpha,\beta
\rightarrow 0$, as expected. We are now ready to analyze the
excitations and mass counts concerning TMHDG for both allowed
signs of $a$. To avoid needless repetitions, we restrict ourselves
to presenting a summary of the main results in Table 1. The
systems that do not appear in this table are tachyonic, {\it
i.e.}, unphysical.

\begin{table}[h]
\caption{Unitarity analysis of the topologically massive
higher-derivative gravity models} {\begin{tiny}
\begin{center}
\begin{tabular}{cccccc}\hline
$a$ & $b$ & $\frac{3}{2}+4c$ &\begin{tabular}{c} excitations and\\
mass counts \end{tabular}& tachyons & unitarity
\\\hline $- 1$&$>0$&$<0$& \begin{tabular}{c} 2 massive\\
spin-2 normal particles\\ 1 massless spin-2\\ non-propagating
particle \\ 1 massive spin-0 ghost
\end{tabular}
&no one& \begin{tabular}{c}non-unitary
\end{tabular}\\\hline$-1$ &$\frac{-1}{2M^2}<b<0$&$>0$& \begin{tabular}{c} 1 massive\\
spin-2 normal particle\\ 1 massless spin-2\\ non-propagating
particle\\ 1 massive spin-2 ghost
\\ 1 massive spin-0 ghost
\end{tabular}
&no one& \begin{tabular}{c}non-unitary
\end{tabular}\\\hline
 +1&$<0$&$<0$&
\begin{tabular}{c} 2 massive spin-2 ghosts\\ 1
massless spin-2 \\ non-propagating particle \\ 1  massive\\ spin-0
normal particle
\end{tabular}
&no one&\begin{tabular}{c}non-unitary
\end{tabular}\\\hline
$+1$ &$0<b<\frac{1}{2M^2}$&$>0$& \begin{tabular}{c} 1 massive\\
spin-2 normal particle\\ 1 massless spin-2\\ non-propagating
particle\\ 1 massive spin-2 ghost
\\ 1 massive spin-0\\ normal particle
\end{tabular}
&no one& \begin{tabular}{c}non-unitary
\end{tabular}\\\hline
\end{tabular}
\end{center}
\end{tiny}}
\end{table}

 \noindent In conclusion, we consider  TMHDG with $a=0$. In this case,

 \begin{eqnarray}
SP_\mathrm{TMDHDG} =\frac{M^2b}{2} \left[ - \frac{T^{\mu\nu}
T_{\mu\nu} -\frac{1}{2}T^2}{k^2} + \frac{T^{\mu\nu} T_{\mu\nu}
-\frac{1}{2}T^2}{k^2 - \frac{4}{M^2 b^2}} \right]
+\frac{1}{b(\frac{3}{2} + 4c)} \frac{\frac{1}{2} T^2}{k^4}.
\nonumber
\end{eqnarray}

\noindent The pole of order two at $k^2=0$ indicates that these
models are unphysical.

\subsubsection{Discussion}
As intuitively expected, TMHDG is non-unitary for $a=\pm 1$;
nonetheless, these models are in general non-tachyonic which means
that  under certain circumstances they may be viewed as effective
field models. Our aim here is to investigate, in passing, the
novel features of one of these non-unitary gauge-invariant
three-term effective field models. To be more specific, we fix our
attention on the first model of Table 1, {\it i.e.}, TMHDG with
$a=-1$, $b>0$, and $\frac{3}{2} + 4c <0$ Ref. 13. We have chosen
the $a=-1$ system because it reduces, in the absence of the
topologically massive term, to higher-derivative gravity with
$a=-1$---an effectively multimass model of the fourth-derivative
order with interesting properties of its own.\cite{14} Now, it can
be shown that the effective non-relativistic potential for the
interaction of two scalars bosons in the framework of  TMHDG with
$a=-1$, $b>0$, and $\frac{3}{2} + 4c <0$, is given by\cite{15}

\begin{eqnarray}
V(r)= 2{ \bar m}^2 {\bar G} \left[ K_0 (rm) - \frac{ K_0
(rM_{+})}{1 + \frac{bM^2_{+}}{2}} - \frac{ K_0 (rM_{-})}{1 +
\frac{bM^2_{-}}{2}}
 \right],
\end{eqnarray}

\noindent where ${\bar m}$ is the mass of one of the neutral
bosons , ${\bar G} \equiv \frac{\kappa^2}{32\pi}$, and $$M_{\pm} =
\frac{1}{bM} \left[ \sqrt{1 +2bM^2} \pm 1 \right].$$ Note that
$V(r)$ behaves as $2{\bar m}^2 {\bar G} \ln \left( \frac{M^{1 +
\frac{b M^2_{+}}{2}}_{+} M^{1 + \frac{b M^2_{-}}{2}}_{-}}{m}
\right)$ at the origin and as $$ 2{ \bar m}^2 {\bar G} \left[
\sqrt{\frac { \pi}{2rm}}\; e^{-rm} - \frac{1}{1 +
\frac{bM^2_{+}}{2}} \sqrt{\frac{\pi}{2rM_{+} }} \; e^{-rM_{+} } -
\frac{1}{1 + \frac{bM^2_{-}}{2}}  \sqrt{\frac{\pi}{2rM_- }}\;
e^{-rM_{-} } \right].$$

\noindent asymptotically. Accordingly, $V(r)$ is finite at the
origin and zero at infinity. The derivative of this potential with
respect to $r$ is in turn given by

\begin{eqnarray}
\frac{dV}{dr} =2{\bar m}^2 {\bar G} \left[ - m K_1(rm) +
\frac{M_{+}}{1 +\frac{bM^2_{+}}{2}} K_1(rM_{+})   + \frac{M_{-}}{1
+\frac{bM^2_{-}}{2}} K_1(rM_{-}) \right]
\end{eqnarray}

\noindent On the other hand, it was shown recently that in four
dimensions
 the propagation of photons in the context of higher-derivative gravity (HDG) is dispersive.\cite{16}
 In other words, gravitational rainbows and semiclassical
HDG can coexist without conflict. On the basis of the fact that
the rainbow effect is currently undetectable, it is possible to
show that $|\beta| \leq 10^{60}$ Ref. 17. How reliable is this
result? The aforementioned constraint is of the same order as that
obtained by testing the gravitational inverse-square law in the
submillimeter regime.\cite{18} Thence, we assume  $b \gg 1$. As a
consequence, Eq. (16) reduces to $$ \frac{dV}{dr} \sim 2{\bar m}^2
{\bar G} \left[ - m K_1(rm) +
 \sqrt{ \frac{2}{b}} K_1\left(r \sqrt{\frac{2}{b}}\right) \right],$$

\noindent implying that the potential $V(r)$, which in this
approximation may be expressed as

\begin{eqnarray}
 V(r) \sim2 {\bar m}^2 {\bar G} \left[ K_0(rm) - K_0\left(r
\sqrt{\frac{2}{b}}\right) \right],
\end{eqnarray}

\noindent is everywhere attractive if $\sqrt{\frac{2}{b}} > m$, is
repulsive if $ m > \sqrt{\frac{2}{b}}$, and vanishes if $m =
\sqrt{ \frac {2}{b}}$. If we appeal to the usual tools of
Einstein's geometrical theory, we arrive at the same conclusions.
In fact, in the weak field approximation the gravitational
acceleration , $\gamma^l = \frac{dv^l}{dt}$, of a slowly moving
test particle is given by $\gamma^l = -\kappa \left[
\frac{\partial}{\partial t} h^l_0 - \frac{1}{2}
\frac{\partial}{\partial l} h_{00} \right]$, which for
time-independent fields reduces to $\gamma^l = \frac{\kappa}{2}
\frac{\partial}{\partial l} h_{00}$. Now, taking into account that
$h_{00} = \frac{2V}{{\bar m} \kappa}$, we obtain $$ \gamma^l =
2{\bar m} {\bar G} \frac{x^l}{r} \left[ -m K_1(rm) +
\sqrt{\frac{2}{b}} K_1 \left(r\sqrt{ \frac{2}{b}} \right)
\right].$$

\noindent Therefore, the gravitational force exerted on the
particle, $$F^l = 2 {\bar m}^2 {\bar G} \frac{x^l}{r} \left[ -m
K_1(rm) + \sqrt{\frac{2}{b}} K_1 \left(r\sqrt{ \frac{2}{b}}
\right) \right], $$

\noindent is everywhere attractive if $\sqrt{\frac{2}{b}} > m$, is
repulsive if $ m > \sqrt{\frac{2}{b}}$ (antigravity), and vanishes
if $m = \sqrt{ \frac {2}{b}}$ (gravitational shielding). It is
remarkable that this force does not exist in general relativity.
It is peculiar to topologically massive higher-derivative gravity.

\section{Final remarks}

We have shown that topologically massive terms cannot be used as a
panacea for curing the non-unitarity of massive
electromagnetic/gravitational models. In truth, the addition of a
Chern-Simons term to a massive electromagnetic/gravitational model
is physically sound only and if only the resulting three-term
system reduces to well-behaved physical models in the suitable
limits. A direct consequence of this fact is that we will never be
able to construct an unitary, massive, topologically massive,
gravitational model. Indeed, the fancy way Einstein-Chern-Simons
theory is built, {\it i.e.}, with the Einstein's term with the
opposite sign, precludes the existence of ghost-free, massive,
topologically massive, gravitational models. Therefore, from a
conceptual point of view, the addition of a topologically massive
term to a massive gravitational model is a complete nonsense: On
the one hand, it does not cure the non-unitarity of the original
model; on the other hand, it spoils the unitarity of admittedly
unitary models. An interesting and elucidatory example is
furnished by $ R + R^2$ gravity, which is defined by the
Lagrangian $\mathcal{L} = \left[ a \frac{2R}{\kappa^2} +
\frac{\alpha}{2} R^2 \right] \sqrt{g}$, with $a= \pm 1$. If
$\alpha >0$, this theory is non-tachyonic regardless of the sign
of $a$; in addition,  it is unitary if $a=+1$, and non-unitary if
$a=-1$. Incidentally, $ R + R^2$ gravity with $a=+1$ is the only
known gravity theory with higher-derivatives that is unitary.
However, topologically massive $ R + R^2$ gravity is non-unitary
for both possible sign choices of $a$ Ref. 19. Yet, a new and
surprising physics emerges when we analyze the  three-term
effective field models that are both gauge-invariant and
non-unitary. In the framework of the electromagnetic models, an
attractive interaction between equal charge particles can be
produced that leads to an unusual planar dynamics: scalar pairs
can condense into bound states. In the framework of the gravity
systems, in turn, unlike what occurs in the context of the insipid
and odorless three-dimensional Einstein's general relativity, we
have a gravitational interaction that can be both attractive and
repulsive. We can also have a null gravitational interaction, such
as in three-dimensional gravity that is trivial outside the
sources. Certainly, these effective field models deserve to be
both much better known and further investigated.

\section*{Acknowledgments}
\noindent  We are very grateful to Prof. S. Deser  for calling our
attention to Ref. 6. A. Accioly thanks CNPq-Brazil for partial
support while M. Dias is indebted  to CAPES-Brazil for full
support.

\end{document}